\documentstyle[epsfig,12pt]{article}  
\newcommand{\bce}{\begin{center}}
\newcommand{\ece}{\end{center}}
\newcommand{\beq}{\begin{equation}}
\newcommand{\eeq}{\end{equation}}
\newcommand{\bea}{\vspace{0.25cm}\begin{eqnarray}}
\newcommand{\eea}{\end{eqnarray}}

\newcommand{\ba}{\begin{array}}
\newcommand{\ea}{\end{array}}

\newcommand{\r}{\mbox{{\boldmath
$\rho$}}}

\newcommand{\qb}{\mbox{{\bf
q}}}

\newcommand{\bb}{\mbox{{\bf
b}}}
\newcommand{\rb}{\mbox{{\bf
r}}}

\newcommand{\pb}{\mbox{{\bf
p}}}

\def\lsim{\mathrel{\rlap{\lower4pt\hbox{\hskip1pt$\sim$}}
    \raise1pt\hbox{$<$}}}         
\def\gsim{\mathrel{\rlap{\lower4pt\hbox{\hskip1pt$\sim$}}
    \raise1pt\hbox{$>$}}}         

    \def\Pom{{\bf
    I\!P}}
    \def\beq{\begin{equation}}
    \def\endeq{\end{equation}}
    \def\bea{\begin{eqnarray}}
    \def\arr{\begin{eqnarray}}
    \def\eea{\end{eqnarray}}

\def\qbq{$q
\bar{q}~$}
\def\q2{$Q^{2}$}
\def\s2{2$S$}

\begin{document}
\thispagestyle{empty}
\vspace*{-2cm}
\begin{flushright}
{\bf\large
FZJ-IKP(Th)-1999-26\\}
\end{flushright}
 
\bigskip
\begin{abstract}
  
\end{abstract}

\begin{center}

  {\large\bf
NUCLEAR-MEDIUM MODIFICATION OF THE $\rho^{0}(1S)$- 
AND $\rho{'}(2S)$-MESONS IN 
COHERENT PHOTO- AND ELECTROPRODUCTION: COUPLED CHANNEL ANALYSIS 
\\
\vspace{1.5cm}
  }
\medskip

{\bf \large N.N. Nikolaev$^{1}$, J. Speth$^{1}$
and B.G. Zakharov$^{2}$
\bigskip\\}
{\sl 

$^{1}$Institut  f\"ur Kernphysik, Forschungszentrum J\"ulich,\\
        D-52425 J\"ulich, Germany\medskip\\
$^{2}$Landau Institute for Theoretical Physics,
        GSP-1, 117940,\\ Kosygina Str. 2, 117334 Moscow, Russia
}
\vspace{4cm}

  {\bf Abstract}
\end{center}

We study medium modifications of the  the $e^{+}e^{-}$ and/or $\mu^{+}\mu^{-}$
mass spectrum in coherent photo- and electroproduction of the $\rho^{0}(1S)$- and 
$\rho'(2S)$-meson resonances on nuclear targets. The analysis is performed within 
the coupled $\rho^{0}(1S),\rho'(2S),...$ channel formalism in which nuclear modifications
derive from the off-diagonal  rescatterings. We find that the effect of
off-diagonal rescatterings on the shape of the $e^{+}e^{-}$ 
mass spectrum in the $\rho^{0}(1S)$-meson mass region is only marginal
but is very important in the $\rho'(2S)$ mass region. The main off-diagonal 
contribution in the $\rho'(2S)$ mass region comes
from the sequential mechanism 
$\gamma^{*}\rightarrow \rho^{0}(1S)\rightarrow \rho'(2S)$, which 
dominates the $\rho'(2S)$ production for heavy nuclei.
 Our results show also that
in the $\rho'(2S)$ mass region there is a considerable effect of  
the interference of the Breit-Wigner
tail of $\rho^{0}(1S)$-meson with the $\rho'(2S)$-meson.

\newpage

\section{Introduction}

The in-medium modification of hadrons in the cold nuclear matter and 
hot QCD medium has been under active investigation during the last years.
In particular, there was considerable interest in the medium effects
for light vector mesons. In several papers the masses of
vector mesons at rest in nuclear matter have been calculated within 
different approaches \cite{BM,Hos,BR,HL,A,Koike,HLS}.
The in-medium effects for moving vector mesons have been discussed 
in refs. \cite{I1,I2,K1,K2,K3}. Recall that in optics medium effects 
are described by the refraction index which for the dilute media is 
calculable in terms of the photon-atom forward scattering amplitude.
Extension of this formalism to fast vector mesons with the 
wavelength $\lambda=1/p$ much shorter than  
separation of nucleons in the nuclear matter gives the in-medium 
mass shift and collisional broadening the form (\cite{I1}, for
the related early works see \cite{K4,Bugg})
\beq
\Delta m_{V}(E)=-2\pi\frac{n_{A}}{m_{V}}\mbox{Re} f(E)\,\,,\,\,\,\,\,\,\,\,
\Delta \Gamma_{V} (E)=\frac{n_{A}}{m_{V}}p\sigma(E)\,\,,
\label{eq:1}
\eeq
where $E$ and $m_{V}$ are the meson energy and in-vacuum mass, $f(E)$ 
is vector meson-nucleon forward scattering amplitude 
and $n_{A}$ is the nuclear matter density.
These formulas are quite general and must hold for any particle in an 
infinite medium if the inelastic rescatterings and/or coupled-channel
effects can be neglected, see below. An experimental observation of the in-medium 
mass shift and collisional broadening  (1) would 
be very interesting.
Potentially it could give information on the $V N$ scattering amplitude
which cannot be measured directly.

In order for the mass shift and collisional broadening (1) to be
observed experimentally 
the typical decay length $L_{d}\sim E/m_{V}\Gamma_{V}$ must be
smaller than the nucleus radius $R_{A}$, i.e., the momentum of the vector 
meson must be smaller than  
\beq
p_{\rho} <6~{\rm GeV},~~ p_{\omega}< 300~ {\rm MeV},~~ p_{\phi}<~200~{\rm MeV}\, ,
\label{eq:2}
\eeq
for the $\rho^{0}$-, $\omega^{0}$-, and $\phi^{0}$-meson, respectively \cite{I1}.  
Thus only for the $\rho^{0}$-meson \footnote{Hereafter wherever appropriate the
$\rho^{0}$ will stand for the ground state $\rho^{0}(1S)$ meson.}
there is a sufficiently broad energy interval
where the relations (1) could be used. 

The applicability limits (\ref{eq:2}) are purely kinematical ones and do not take 
into account possible interference of decays of the vector meson 
inside and outside the target nucleus and the quantum effects in
 production of the vector mesons.   
The both effects were studied, and found to be important, in the
recent \cite{K1} Glauber-Gribov 
multiple-scattering theory \cite{Glauber,Gribov} analysis  
of coherent $\rho^{0}$-meson photoproduction
observed in $e^{+}e^{-}$ and/or $\mu^+\mu^-$ dilepton mode. The dilepton mass 
spectrum  was shown to have a two-component structure corresponding to 
the decays of $\rho^{0}$-meson inside and outside the target nucleus.
The inside component can be approximately described by a Breit-Wigner formula
with the in-medium modified mass and width as predicted by Eqs. (1). However,
since the nucleus has a finite size, this component with a broad width 
does not develop a genuine pole in the complex plane of the invariant mass, $M$, 
of the lepton pair.
The genuine pole in the complex $M$-plane pole comes only from 
the $\rho^{0}$-meson decays in vacuum and this outside component can be described
by the standard Breit-Wigner formula with the in-vacuum mass and width.
Ref. \cite{K1} found that even at 
low energy $E\sim 2$ GeV such that $R_{A}/L_{d}\sim 2$ for heavy nuclei, 
the interference between the inside and
outside components is substantial and produces a complex dilepton
mass spectrum which cannot be described in terms of a Breit-Wigner
formula with a definite mass and width. For instance, at $E=2$ GeV the 
dilepton  mass spectrum was found to develop a minimum near the 
$\rho^{0}$-meson mass. An experimental observation of this phenomenon
would be of great interest. On the theoretical side, this calls for
a numerical analysis within tested models of photoproduction on how
a model analysis of such data 
would allow to distinguish the inside and outside components and 
extract the $\rho^{0}$-meson mass shift and collisional broadening  
for the inside 
component.

In the present paper we study the coherent
reaction $\gamma^{*} A\rightarrow V A \rightarrow 
e^{+}e^{-}A, \mu^+\mu^- A$, where $\gamma^{*}$ is a 
real or virtual photon and the target nucleus remains in the ground state,
within the coupled-channel approach extending 
over the early work \cite{K1} in which the numerical 
calculations have been performed neglecting the off-diagonal rescatterings of
$\rho^{0}$-meson. 
The coupled-channel analysis presented here is based on our well tested
colour dipole 
approach (see, for instance, \cite{NNPZ1} and references therein), which 
was earlier successfully used in the analysis of the 
data on $\rho^{0}$ and $J/\Psi$ 
electroproduction on nuclear targets at high energies
\cite{BKMNZ,KNNZ1,KNNZ2} and vector meson production at HERA 
\cite{NNPZ1,NNPZ2,NNPZZ}.
This analysis is of interest for two reasons. First,
including the off-diagonal rescatterings allows one to check the accuracy 
of the one-channel approximation in the $\rho^{0}(1S)$-meson mass region
$0.5< M < 1$ GeV studied in ref. \cite{K1}. 
On the other hand, in the coupled-channel approach one can extend
the mass region and investigate 
the medium effects for the 2$S$ state $\rho'(2S)$ for which the sequential
off-diagonal mechanism $\gamma^{*}\rightarrow \rho^{0}(1S)\rightarrow \rho'(2S)$ 
is potentially important. This extension to the $\rho'(2S)$-meson 
is of great interest in itself. 
The key feature of  photoproduction of the $2S$ vector mesons
on a free nucleon is strong suppression due to the nodal structure of the 
wave function of the 2$S$ state \cite{Jpsi,BKMNZ}. In ref. \cite{NNZ1} it was shown 
that the node 
effect can lead to an anomalous $A$- and $Q^{2}$-dependence of $\rho'(2S)$ photo- and 
electroproduction.
This effect may help to resolve the long standing problem of the 
$D$-wave vs. $2S$-wave assignment for the $\rho'(1480)$ and 
$\rho'(1700)$ states.
The strong suppression of the cross section for the
$\rho'(2S)$-meson
as compared to the  $\rho^{0}(1S)$-meson makes the experimental study of this phenomenon 
a challenging task. In particular, the mass spectra of the final particles
in the $\rho'(2S)$ mass region can be affected by the interference
with the $\rho^{0}(1S)$-meson Breit-Wigner tail. 
Our colour dipole coupled-channel approach describes very well the suppression
of $\Psi'$ production as compared with $J/\Psi$
observed by the NMC \cite{NMCJpsi,NMCPsi'}, E687 \cite{E687} collaborations
and H1 collaboration \cite{H1} and provides a sound framework for
understanding the the prospect of experimental 
study of the $\rho'(2S)$-meson production $\gamma^{*} A\rightarrow 
\rho^{0}(1S)A, \rho'(2S)A \rightarrow e^{+}e^{-}A$. In the reported numerical analysis we focus on
the range of energies of the forthcoming high-luminosity 
experiments at TJNAF .

The paper is organized as follows. In section 2 we give the basic
formulas of the coupled channel approach to the 
coherent $\gamma^{*}A\rightarrow e^{+}e^{-}A, \mu^{+}\mu^{-}A$ reaction.
In section 3 we discuss evaluation of the diffraction scattering
matrix.
The numerical results on nuclear modifications of the in-nucleus decay
component of the dilepton mass spectrum 
are presented in section 4..
The results are summarized in section 5.

\section{The coupled channel formalism}

The coupled-channel formalism for the $\gamma A\rightarrow e^{+}e^{-}A$
reaction is found in ref. \cite{K1}, see also our early work \cite{VMJETP}. 
For this reason,  here we discuss it briefly and give only the basic formulas
which are necessary for understanding technical aspects of our approach.

The standard Glauber-Gribov multiple-scattering
theory \cite{Glauber, Gribov} for coherent interaction of the 
projectile of energy E with a nucleus is equivalent to solving the set 
of coupled-channel eikonal  
wave equations \cite{K2}:
\beq
\left [-\frac{\partial^{2}}{\partial z^{2}} +\hat m^{2} +\hat U(\rb)
\right]_{ij}\Psi_{j}(\rb)=E^{2}\Psi_{i}(\rb)\,.
\label{eq:3}
\eeq
Here the $z$-axes is chosen along the photon momentum;  
 $\Psi_{i}$ is the wave function for the channel $|i\rangle$ 
which can be a hadronic resonance or the initial photon;
$\hat m^{2}$ is the diagonal mass operator with eigenvalues 
$m_{i,}^{2}=(m_{i}^{2}-im_{i}\Gamma_{i})\delta_{ij}$,
where $m_{i}$ and $\Gamma_{i}$ are the  in-vacuum mass and 
width of the state $|i\rangle$;,
for the incident photon $\Gamma_{\gamma^{*}}=0$ and 
$m_{\gamma^{*}}^{2}=-Q^{2}$ where $Q^{2}$ is the photon virtuality
and the optical potential 
$\hat U(\rb)$ in (\ref{eq:3}) equals 
\beq
U_{ij}(\rb)=-4\pi\langle i|\hat f|j\rangle n_{A}(\rb)\,,
\label{eq:4}
\eeq
where $n_{A}(\rb)$ is the nuclear number density and $\hat f$ is 
the forward scattering matrix in the normalization
$$
\mbox{Im}\langle i|\hat f|i\rangle={p_{i}\over 4\pi}\sigma_{tot}(iN\rightarrow iN)\, .
$$
The boundary condition on the front face of the nucleus,
$z=-R_{A}$, is $|i\rangle = \delta_{i\gamma^{*}}$\, .
 
The probability amplitude for the coherent transition
$\gamma^{*}A\rightarrow e^{+}e^{-} A$ be written
in terms of the solution $\Psi_{i}(\rb)$ of eq.~(\ref{eq:3}) as \cite{K1}
\beq
T(E,M,\pb_{\perp})=N\sum\limits_{i=h}
\langle e^{+}e^{-}|t|i\rangle\int d^{2}\bb dz
\exp[-i(p_{z}z+\pb_{\perp}\bb)]\Psi_{i}(\rb)\,,
\label{eq:5}
\eeq
where $\rb=(\bb,z)$, $M$ is the invariant mass of the $e^{+}e^{-}$ pair,
$p_{z}$ and $\pb_{\perp}$ are its longitudinal and transverse momenta, 
respectively, $\langle e^{+}e^{-}|t|i\rangle$ is the probability amplitude
for $i\rightarrow e^{+}e^{-}$ transition, $N$ is a normalization factor
which is immaterial from the point of view of the shape of 
the $e^{+}e^{-}$ mass spectrum.
The summation in (\ref{eq:5}) goes only over the hadronic states,
which in our case are vector mesons $\rho^{0}(1S),\rho'(2S)$ etc. 
For a heavy nucleus with a mass number $A\gg 1$ and at $E\gg M$ the
nuclear recoil can be neglected and 
the longitudinal momentum of the $e^{+}e^{-}$ pair equals
\beq
p_{z}\approx E-\frac{Q^{2}+M^{2}+\pb_{\perp}^{2}}{2E}\,\,.
\label{eq:6}
\eeq
In the coherent production $\pb_{\perp}^{2} \lsim 1/R_{A}^{2}$ 
and in what follows we focus on $\pb_{\perp}=0$ and suppress this argument.
Then by virtue of (\ref{eq:6}) the $p_{z}$-dependence of 
the spatial integral in the right hand side of
Eq. (\ref{eq:5}) transforms directly into the $M$-dependence of the amplitude
$T(E,M,\pb_{\perp})$ and nuclear modification of the $e^{+}e^{-}$
spectrum. 

At high energies the solution of (\ref{eq:3})
for the hadronic sector 
which we need for evaluation of the amplitude (\ref{eq:5}) 
can be written in the form
\beq
\Psi_{h}(\bb,z)=\langle h|\hat S(\bb,z)|\gamma^{*}\rangle
\exp(ip_{\gamma^{*}}z)\,,
\label{eq:7}
\eeq
where the operator $\hat S$ is given by
\beq
\hat{S}(\bb,z)=
\hat{P}_{z}\exp\left\{-\frac{i}{2p_{\gamma^{*}}}
\int\limits^{z}_{-\infty}d \xi
\left[\hat m^{2}+Q^{2}+\hat U(\bb,\xi)\right]
\right\}\,.
\label{eq:8}
\eeq
Here, $\hat P_{z}$ is $z$-ordering operator, and $p_{\gamma^{*}}$ is the
photon momentum.
For the numerical calculations it is convenient to treat in
Eq. (\ref{eq:8}) the
off-diagonal part of the optical potential in the hadronic sector 
as a perturbation while the diagonal transitions are included to
all orders.
Then, following Ref. \cite{VMJETP}, one can represent the matrix 
elements of the operator
$\hat{S}(\bb,z)$
in the form of the $\nu$-fold off-diagonal rescatterings
series
\beq
\langle h|\hat{S}(\bb,z)|\gamma^{*}\rangle=
\sum\limits_{\nu=0}^{\infty}
\langle h|\hat{S}^{(\nu)}(\bb,z)|\gamma^{*}\rangle\,,
\label{eq:9}
\eeq
where
\beq
\langle h|\hat{S}^{(0)}(\bb,z)|\gamma^{*}\rangle=
-\frac{1}{2}
\sigma_{h\gamma^{*}}
\int\limits^{z}_{-\infty}dz_{1} n_{A}(\bb,z_{1})
\exp[i k_{\gamma^{*}h}(z-z_{1})-{1\over 2}
t(\bb,z,z_{1})\sigma_{hh}]\,,
\label{eq:10}
\eeq
\bea
\langle h|\hat{S}^{(\nu)}(\bb,z)|\gamma^{*}\rangle=
\left(-\frac{1}{2}\right)^{\nu+1}
\sum\limits_{i_{1},...i_{\nu}}\sigma^{'}_{hi_{\nu}}
\sigma^{'}_{i_{\nu}i_{\nu-1}}\cdots
\sigma_{i_{1}\gamma^{*}}
\exp(i k_{\gamma^{*}h}z)
\,\,\,\,\,\,\,\,\,\,\nonumber\\ \times
\int\limits^{z}_{-\infty}dz_{\nu+1} n_{A}(\bb,z_{\nu+1})
\exp[i k_{hi_{\nu}}z_{\nu+1}-{1\over 2}
t(\bb,z,z_{\nu+1})\sigma_{hh}]
\,\,\,\,\,\,\nonumber\\ \times
\int\limits^{z_{\nu+1}}_{-\infty}
dz_{\nu} n_{A}(\bb,z_{\nu})
\exp[i k_{i_{\nu}i_{\nu-1}}z_{\nu}-{1\over 2}
t(\bb,z_{\nu+1},z_{\nu})\sigma_{i_{\nu}i_{\nu}}]
\cdots
\,\,\nonumber\\
\times
\int\limits^{z_{2}}_{-\infty}
dz_{1}
n_{A}(\bb,z_{1})
\exp[i k_{i_{1}\gamma^{*}}z_{1}-{1\over 2}
t(\bb,z_{2},z_{1})
\sigma_{i_{1}i_{1}}]\,, \;\;\;\,\,\,\,\,\,\,\,\nu\ge 1\,.
\label{eq:11}
\eea
Here, $\sigma^{'}_{ik}=\sigma_{ik}-\delta_{ik}\sigma_{ii}$,
the matrix $\hat{\sigma}$
is connected with the forward diffraction
scattering matrix 
\beq
\hat{f}={ip_{\gamma^{*}}\over 4\pi}\hat{\sigma}\,, 
\label{eq:12}
\eeq 
\beq
t(\bb,z_{2},z_{1})=\int_{z_{1}}^{z_{2}}dz n_{A}(\bb,z)\,
\label{eq:13}
\eeq
is the partial optical thickness, and 
\bea
k_{ij}=\frac{m^{2}_{i}-im_{i}\Gamma_{i}-m^{2}_{j}+im_{j}\Gamma_{j}}{2E}\,, \\
k_{h\gamma^{*}}=-k_{\gamma^{*}h}=\frac{m^{2}_{i}-im_{i}\Gamma_{i}+Q^{2}}{2E}\,.
\label{eq:15}
\eea
The exponential factor $\exp[i k_{\gamma^{*}h}(z-z_{1})-{1\over 2}
t(\bb,z,z_{1})\sigma_{ii}]$ in (\ref{eq:10}), (\ref{eq:11}) sums 
elastic, diagonal,  $i N$ rescatterings to all orders.

The real part of (\ref{eq:15})  - the longitudinal momentum transfer
in $j\rightarrow i$ transition - rises with 
the difference between $m_{i}$ and $m_{j}$. Consequently, 
the oscillating exponential factors in (\ref{eq:10}), (\ref{eq:11}))
lead to the form factor suppression of the contribution 
from heavy intermediate resonance states and the related form
factor suppression of heavy mass production and of the coherent 
cross section at large $Q^{2}$, when the
longitudinal momentum transfer $k_{h\gamma^{*}}$ in the 
$\gamma^{*}\to h$ transition becomes large. Precisely the same form 
factor effect
generates strong dependence of the probability amplitude (\ref{eq:5}) 
on the mass of the $e^{+}e^{-}$ pair so that the shape of resonances 
would differ strongly from the standard Breit-Wigner one.
 
Eq.~(\ref{eq:5}) quantifies the separation of the production amplitude
$T(E,M)$ into the inside and outside components. Namely, beyond the target
nucleus the $z$-dependence of wave functions $\Psi_{i}(z,\bb)$ follows
the in-vacuum decay law and the corresponding outside contribution to 
$T(E,M)$ has the familiar form of the sum of Breit-Wigner amplitudes with residues 
proportional
to $\Psi_{i}(z=+R_{A},\bb)$. The nuclear effects cam modify dramatically
both the relative amplitude and phase of these residues compared to
$\sigma_{h\gamma^{*}}$ for the free nucleon case. As explained in the
Introduction, the inside contribution is a Fourier transform over the 
finite range $-R_{A}< z < R_{A}$ and could develop the Breit-Wigner form 
with the shifted mass and collision-broadened width only provided that
$L_{d} \ll R_{A}$.

The first order term (\ref{eq:10}) corresponds to the standard Glauber 
approximation
when the state $|h\rangle$ produced in the $\gamma^{*}\rightarrow h$
transition then propagates through the nucleus without inelastic
rescatterings. 
The correction from the off-diagonal rescatterings 
is given by (\ref{eq:11}) and is responsible for the color 
transparency phenomenon in electroproduction
of vector mesons at high $Q^{2}$
where it changes drastically the cross section as compared with the Glauber
model predictions.
However, as will be seen from our results for 
$\gamma^{*}A\rightarrow \rho'(2S)A$ reaction
the off-diagonal effects come into play already at moderate values of the 
photon virtuality $Q^{2}\lsim 1$ GeV$^{2}$. 
This is a consequence of strong suppression of the direct 
$\gamma^{*}\rightarrow \rho'(2S)$
transition. As a result, the sequential mechanism 
$\gamma^{*}\rightarrow \rho^{0}(1S)\rightarrow \rho'(2S)$, involving 
the off-diagonal $\rho^{0}(1S)\rightarrow \rho'(2S)$ rescattering, becomes important 
even at $Q^{2}=0$.

\section{ Calculation of the forward diffraction matrix}

As an input to the coupled-channel calculations 
one needs the forward diffraction matrix.
At high energies it can be written as a sum of the Pomeron and reggeon 
contributions
\beq
\hat \sigma =\hat \sigma_{\Pom}+\hat \sigma_{R}\,.
\label{eq:16}
\eeq
At energies $E\gsim 2$ GeV to be 
considered in our paper, the dominating contribution to the $\hat\sigma$
comes from the Pomeron exchange. 
We evaluate the Pomeron exchange within the 
dipole approach describing the resonances as  nonrelativistic 
$q\bar{q}$ states.
Then the Pomeron contribution to the diffraction matrix element $\sigma_{ij}$
for hadronic states can be written as
\beq
\langle i|\hat{\sigma}_{\Pom}|k\rangle=
\int d^{2}\r dz
\psi_{i}^{*}(\r,z)\sigma(\rho)\psi_{k}(\r,z)\,,
\label{eq:17}
\eeq
where $\r$
is the transverse size of the \qbq pair,
$\psi_{i,k}(\r,z)$ are the
wave functions describing
the \qbq states,
and $\sigma(\rho)$ is the cross
section of interaction of the \qbq pair
with a nucleon.

We also need the excitation matrix elements $\langle i|\hat{\sigma}|\gamma^{*}\rangle$
for the $\gamma^{*}\rightarrow q\bar{q}$ excitation  on a free nucleon
which in the non-relativistic approximation proceed into $q\bar{q}$ states 
with the sum of quark helicities equal to the photon helicity \cite{Jpsi}.
Using the corresponding perturbative light-cone wave function of the 
virtual photon \cite{NZ1}, we
have \cite{Jpsi}
\beq
\langle i|\hat{\sigma}|\gamma^{*}\rangle=C
\int d^{2}\r 
\psi_{i}^{*}(\r,z=0)\sigma(\rho)K_{0}(\epsilon\rho)\,,
\label{eq:18}
\eeq
where
\beq
\epsilon^{2}=m_{q}^{2}+Q^{2}/4\,,
\label{eq:19}
\eeq
$m_{q}$ is the quark mass, $K_{0}(x)$ is the modified
Bessel function. In the present paper we focus on the 
nuclear-modification of the shape of resonances, and 
the absolute value of the normalization factor C in Eq. (\ref{eq:18}), 
and in Eq. (\ref{eq:5}) as well, is
immaterial.

We use the oscillator wave functions for the $q\bar{q}$ states, which
simplifies considerably the numerical calculations: because of
azimuthal symmetry of $\sigma_{\Pom}(\rho)$ 
only the off-diagonal rescatterings change only the  
transverse excitations with zero azimuthal quantum number
can be excited in the intermediate state. 
Excitation energy of the transverse $q\bar{q}$ oscillator equals 
$2\hbar\omega$, here $\omega$ is the 
oscillator frequency. We take in our analysis the quark mass  
$m_{q}=m_{\rho}/2$ and the oscillator frequency
$\omega=(m_{\rho'}-m_{\rho^{0}})/2\approx 0.35$ GeV , assuming 1480 MeV
for the mass of the radial excitation of $\rho^{0}$-meson.
For the widths of the first two excitations we use the values 
$\Gamma_{\rho^{0}}\approx 150$ MeV
and $\Gamma_{\rho'}\approx 285$ MeV and, following the string 
model \cite{Nussinov}, assume $\Gamma_{i}\propto m_{i}$ for higher
lying states. We are not sensitive to this latter assumption, though,
because the contribution of the higher excitations turns out to be very small. 

The experimental data on the low-$x$ structure function $F_{2}$ and vector meson
electroproduction off nucleon can be described by
representing $\sigma (\rho)$ as a sum of the energy
dependent perturbative and energy independent nonperturbative
components \cite{NNPZ1,NNPZZ,BFKLhera}. Here we focus on the
energy region $E\lsim 20$ GeV where the energy dependence of
the dipole cross section can be neglected to first approximation.
Also, in the present analysis we are sensitive to $\sigma(\rho)$
mostly in the nonperturbative region of $\rho \gsim 0.5$. 
Here the gross features of $\sigma(\rho)$ are well parameterized by
the two-gluon exchange model of the Pomeron \cite{DGM1,DGM2}:
\beq
\sigma(\rho)=\frac{16\alpha_{S}^{2}}{3}
\int d^{2}\qb\,\frac{
[1-\exp(i\qb\r)]
[1-G_{2}(\qb,-\qb)]}
{(\qb^{2}+\mu_{g}^{2})^{2}}\,\,,
\label{eq:20}
\eeq
here $G_{2}(\qb_{1},\qb_{2})=\langle N|\exp(i\qb_{1}\rb_{1}+i\qb_{2}\rb_{2})
|N\rangle$
is the two-quark form factor of the nucleon, $\mu_{g}0.3$ GeV is
an infrared cutoff. 
It reproduces the colour transparency property $\sigma(\rho)\propto \rho^{2}$
at small $\rho$, by which  the point-like $q\bar{q}$ system,
which can be represented as a superposition of an infinite set of the 
resonance states, propagates through the nucleus without interaction. 
This color transparency phenomenon in terms of the resonance 
states is connected with exact cancellation
of the diagonal and off-diagonal amplitudes \cite{NCT}.
The coupling constant $\alpha_{S}$
was normalized so the Pomeron contribution to the $\rho^{0} N$ total cross section 
$\sigma_{\Pom}^{tot}(\rho^{0} N)= \langle \rho^{0}|\hat\sigma_{\Pom}|\rho^{0}\rangle 
\approx 20$ mb. In this case the two-gluon formula 
gives at $\rho\gsim 0.5$ fm the $\sigma(\rho)$ which is close to
the dipole cross section extracted from the analysis of the experimental
data on electroproduction of vector mesons \cite{NNPZ1}.  Hereafter
we shall consider real, $Q^{2}=0$, and virtual, $Q^{2}=1$ GeV$^{2}$,
photoproduction. With our nonrelativistic wave functions we find
\begin{equation}
R(2S/1S)={\langle \rho'(2S)|\hat{\sigma}|\gamma^{*}\rangle \over
\langle \rho^{0}(1S)|\hat{\sigma}|\gamma^{*}\rangle}=\left\{
\begin{array}{rl}  
0.21, & ~~~Q^{2}=0 ,\\[3mm]
0.5, & ~~~Q^{2}=1~{\rm GeV}^{2}\,,
\end{array}
\right.
\label{eq:21}
\end{equation}
which is close to 
predictions of ref. \cite{NNPZ1} obtained for the relativized 
wave functions. 
At $\rho\lsim 0.5$ fm the present parameterization gives $\sigma(\rho)$
somewhat larger than that of ref. \cite{NNPZ1}. However, because of
the larger quark mass in the present nonrelativistic model for
$q\bar{q}$ states the resulting diffractive matrix  
turn out to be close to that of ref. \cite{NNPZ1}.
At this point, it must be made clear that the two-gluon parameterization 
(\ref{eq:20}) of $\sigma(\rho)$ is oriented towards 
description of the combined nonperturbative+perturbative 
dipole cross section and $\mu_{g}$ is a phenomenological
parameter which must not be taken at the face value.
The analysis \cite{BFKLhera} of low-$x$ HERA data on the proton
structure function $F_{2p}$
 within the generalized BFKL equation \cite{gBFKL},
and the nonperturbative evaluation of the gluon correlation
radius \cite{Shuryak} yield a clearcut evidence in favor of
the infrared cutoff $\mu_{G} \sim$ 0.75 GeV for the perturbative
dipole cross section.

In our calculations we take into account the first 4 transverse excitations.
The Pomeron contribution to the 
diffraction matrix in 
terms of these transverse oscillator states which we obtained from 
Eq. (\ref{eq:17}) 
for our parameterization (\ref{eq:20}) of $\sigma(\rho)$ 
is given by
\beq
\langle i|\hat\sigma_{\Pom}|k\rangle\approx
\left(\begin{array}{cccc}
20 & -10.1 & -4.8 & 2.4\\ 
-10.1 & 30.2 & -8.0 & 5.2\\
-4.8 & -8.0 & 32.9 & 6.5\\
2.4  &  5.2 & 6.5  & 34.1\\
\end{array}\right)_{ik}\,\,.
\label{eq:22}
\eeq
Here the matrix elements in (\ref{eq:22}) are in units of mb,
$i$ and $k$ are the radial quantum numbers of the transverse
oscillators (as was  said above the Pomeron exchange does not change
the longitudinal quantum number).
The matrix (\ref{eq:22}) shows clearly the decrease of the off-diagonal 
amplitudes with increase of the difference between the initial and final
radial quantum numbers $|i-k|$ which derives from the oscillation of 
the resonance wave function and, in conjunction with the form factor 
effect, suppresses the contribution
of higher excitations in the production amplitude matrix element (\ref{eq:11}).
Using the three-dimensional $\rho'(2S)$-meson wave function from (\ref{eq:22}) 
one obtains for the Pomeron contribution to the 
$\rho'(2S) N$ total cross section  
$\sigma_{\Pom}^{tot}(\rho'(2S) N)=
\langle \rho'(2S)|\hat\sigma_{\Pom}|\rho'(2S)\rangle \approx 27$ mb.

In parameterizing the reggeon contribution to the diffraction matrix
we assume that
the secondary reggeon exchanges can be treated in terms of 
scattering amplitudes for
the quark (or antiquark) making up the $q\bar{q}$
state as predicted by the dual parton model \cite{DPM} 
and quark gluon string model \cite{QGSM} based on the idea of the 
topological expansion \cite{TE}. 
In this case one
can neglect the reggeon contribution to the off-diagonal transitions,
and for all excited $\rho'$ states the reggeon contribution
to diagonal $\rho' N$ scattering amplitudes turn out to be equal to the 
reggeon contribution to
the $\rho^{0}(1S) N$ scattering amplitude. 
This amplitude is dominated by the contribution
of the Regge pole $P'$ which 
can be written in the Regge approach as
\beq
\langle \rho^{0} |\hat \sigma_{R} |\rho^{0}\rangle= 
\r_{P'}\left(\frac{s}{s_{0}}\right)^{\alpha_{P'}-1}
\left[1+i\frac{1+\cos \pi\alpha_{P'}}{\sin \pi\alpha_{P'}}\right]\,,
\label{eq:23}
\eeq
where $s=m_{N}^{2}+2Em_{N}$.
In our analysis we take the standard reggeon intercept
$\alpha_{P'}=0.5$. The residue $r_{P'}$ has been
adjusted to reproduce at $E\sim 10$ GeV the real part of the 
$\rho^{0} N$ scattering 
amplitude extracted in ref. \cite{K3} from the experimental data for 
$\rho^{0}$-meson photoproduction using the vector dominance model.
For $s_{0}=1$ GeV$^{2}$ this gives $r_{P'}\approx 15$ mb.
The nonzero Re/Im ratio for the diffraction scattering matrix leads
to the mass shift for the resonance states decaying 
inside the nucleus.  
For the $\rho^{0}$-meson in the energy region $E\sim 2-20$ GeV considered in 
the present paper it gives 
$\Delta m_{\rho^{0}}\sim $ 50-100 MeV.

For evaluating the $e^{+}e^{-}$ mass spectrum we need also
the transition amplitude
$\langle e^{+}e^{-}|t|i\rangle$ which enters Eq. (\ref{eq:5}). Following
the authors of ref. \cite{K1} we neglect a possible smooth 
$M$-dependence of this transition amplitude as compared with that coming from
the spatial integral in the right hand side of (\ref{eq:5}), and take 
$\langle e^{+}e^{-}|t|i\rangle\propto \psi_{i}(\rb=0)$ as predicted by the 
nonrelativistic model of $q\bar{q}$ states.
Note that in the nonrelativistic approach the $D$-wave $q\bar{q}$ 
state does not
contribute to the di-lepton production. For this reason, the absence of
splitting the $2S$- and $D$-states in the oscillator model is 
not very important
from the point of view of the $e^{+}e^{-}$ mass spectrum.

The applicability of the full fledged couple-channel formalism 
depends on the two space-time scales: the formation length 
$L_{f}$, associated 
with the $i\rightarrow k$ transitions,
\beq
L_{f}={1 \over k_{ik}}=
{2E \over m_{i}^{2}-m_{k}^{2}} \sim {2E \over m_{\rho'}^{2} -m_{\rho^{0}}^{2}}
\approx 0.25 fm \cdot {E \over 1 {\rm GeV}}\, ,
\label{eq:24}
\eeq 
and the coherence length
\beq
L_{c}=
{1 \over k_{h\gamma^{*}}}={2E \over M^{2}+Q^{2}} 
\approx 0.75 fm {E \over 1 {\rm GeV}}\cdot {m_{\rho^{0}}^{2} \over M^{2}+Q^{2}}
\label{eq:25}
\eeq
associated with the transition $\gamma^{*}\rightarrow i$. Strictly speaking,
evaluation of diffraction matrix from equation (\ref{eq:17}) and
of excitation amplitudes from (\ref{eq:18}) in terms of the color dipole 
cross section only holds if $L_{f}> R_{N}$ and $L_{c} > R_{N}$.
For the interesting to us region $Q^{2}\lsim 1$ GeV$^{2}$ and 
for the $\rho^{0}(1S)$- and $\rho'(2S)$-mesons the full fledged coupled-channel 
effects develop only at $E\sim 5-8$ GeV. 
However, at lower energy the only change is the decoupling
of excitation of higher excitations from the photon and of 
off-diagonal diffractive transitions to and from higher excitations 
and we can stretch the formalism even down to $E=2$ GeV. At this
energy we have a single-channel problem with $\gamma^{*} \to
\rho^{0}$ excitation followed by diagonal $\rho^{0} N$ scattering.
For the evaluation of the relevant diagonal elastic
rescatterings we only need the $\rho^{0} N$ total cross section
and the color dipole value of $\langle \rho^{0}|\hat\sigma_{\Pom}|\rho^{0}\rangle$
does still a good job for the almost energy independent Pomeron contribution.
Simultaneously, $\langle \rho^{0}|\hat{\sigma}|\gamma^{*}\rangle$ enters
only as the overall normalization and whether it is evaluated from
eq. (\ref{eq:18}) or within different approach does not affect 
nuclear modifications of the  $e^{+}e^{-}$ mass spectrum.

\section{Numerical results}

\subsection{The input parameters}

We have performed the numerical calculations 
for the energies $E=2,\,5,\,10, 20$ GeV at $Q^{2}=0$ and 1 GeV$^{2}$
for the target nuclei
$^{9}$Be, $^{56}$Fe and $^{207}$Pb.
For the parameterization of the diffractive matrix and of the photon
and vector meson wave functions see section 3. 
For the nuclear matter density in the light target nucleus $^{9}$Be
we use the oscillator shell model with the
oscillator frequency
adjusted to reproduce the experimental value of the root-mean-square
radius of the charge distribution
$\langle r^{2}\rangle^{1/2}_{^{9}Be}=2.51$ fm \cite{Atdata}.
For the target nucleus $^{56}$Fe the parameterization of the
nuclear density by a sum of Gaussians from ref. \cite{Atdata}
was used. For $^{207}$Pb  we use the Wood-Saxon parameterization of
the nuclear density with parameters borrowed from \cite{Atdata}.

\subsection{The basis of vector meson states}

As stated above, in our numerical calculations we include 4 transverse resonance 
states. For the number of the off-diagonal rescatterings of the $q\bar{q}$ state 
we take $\nu=2$.
We checked that in our kinematic region 
the contribution from higher excitations and higher order off-diagonal
rescatterings can safely be neglected.
Furthermore, an approximation of 
the first two states 
and one  off-diagonal rescattering of the $q\bar{q}$ state
is sufficient for all the practical purposes.
Our principal interest is in the interplay of nuclear effects and interference 
of the $\rho^{0}(1S)$ and $\rho'(2S)$ and we did not include the 
numerically smaller contributions from the 
$\omega$- and $\phi$-mesons and their excitations. These long-lived
resonances decay for the most
part outside the target nucleus and the nuclear effects 
do not modify considerably their shapes.

\subsection{The presentation of results}

Due to the above mentioned form factor effect connected with the 
longitudinal momentum transfer and the suppression of the $\rho'(2S)$ 
production by the node effect
the amplitude (\ref{eq:5}) decreases strongly with increase of the mass of 
the $e^{+}e^{-}$ pair. 
In order to facilitate graphical presentation of the results,
following Ref. \cite{K1} we use the 
scaled amplitude
\beq
T'(E,M)=M^{2}T(E,M)/A\,.
\label{eq:26}
\eeq
Since in the absence of absorption effects (\ref{eq:5}) gives
the amplitude $\propto A$, in (\ref{eq:26}) we also introduced 
the factor $1/A$. Our principal numerical results for the 
mass spectrum  are shown in figs. 1-6 in the form of
$$
|T'|^{2} \propto \left. {M^{4} \over A^{2}}{d\sigma \over dM^{2}d p_{\perp}^{2}}
\right|_{p_{\perp}=0}\, . 
$$
We focus on the mass range $M < 1.75$ GeV and include as final states
the $\rho^{0}(1S)$ and $\rho'(2S)$ vector mesons. 

In order to see better the resonance behaviour of production amplitudes
we also show in figs. 7-9 the Argand plots for $T'$.

Now we shall comment on the salient features of the $Q^{2}$- , energy and nuclear
target dependence of these mass spectra.

\subsection{The nucleon target}

The reference $e^+e^-$ mass spectrum for the proton target and the photon 
energy $E=5$ GeV is shown by thick solid curves in figs. 1b-6b.
In the approximation of the energy independent 
dipole cross section $\sigma(\rho)$ it does not depend on the 
photon energy $E$. The mass spectrum exhibits the well separated 
$\rho^{0}(760)$ and $\rho'(1480)$ resonance peaks. Recall the factor
$M^{4}$ which makes the $\rho'(2S)$ tail about flat at large $M^{2}$.

The suppression by the node effect is lifted with increasing $Q^{2}$, 
see eq.~(\ref{eq:21}), 
and the comparison of the mass spectra for real, $Q^{2}=0$, and
virtual, $Q^{2}=1$ GeV$^{2}$, photoproduction shows clearly the 
predicted rise of the $\rho'(2S)/\rho^0(1S)$ ratio with increasing $Q^{2}$
\cite{Jpsi,BKMNZ,NNZ1}. 
This rise of the $\rho'(2S)$ signal with $Q^{2}$ is clearly seen 
from a comparison of the Argand diagrams of fig.~7b and fig.~7b.
 
\subsection{Mass shift {\sl vs.} nuclear form factor effects:
the $\rho^{0}$ region at low energy}

In figs. 1a-6a we show the mass spectrum for low energy, $E=2$ GeV,
at which the $\rho'(2S)$ production is negligible and the 
single-channel Glauber approximation (\ref{eq:10}) holds. 
The energy $E=2$ is very low such that the $\rho^{0}$ would be expected 
to decay within nucleus. Although this reasoning is correct, 
our results show that the effect 
of the mass shift (\ref{eq:1}) due to the real part of the forward 
$\rho^{0} N$ scattering amplitude proves to be marginal in the nuclear
modification of the inside component of the production amplitude. 
The point is that mass shift enters the exponent of 
the integrand in (\ref{eq:10}) via the extra phase 
\beq
\phi= {1\over 2}
t(\bb,z,z_{1}){\rm Im}\sigma_{hh} \approx 
{1\over 2}\cdot r_{\Pom'} n_{A}(z-z_{1})\sqrt{{s_{0} \over s}} 
(z-z_{1}) 
\label{eq:27}
\eeq
which must be compared to the phase $k_{\gamma^{*}h}(z-z_{1})=-(z-z_{1})/L_{c}$ 
from the finite coherence length. Evidently, the significance of the mass shift 
for the inside component of the dilepton production amplitude is
controlled by the parameter
\beq
\eta = {1\over 2}\cdot r_{\Pom'} n_{A}\sqrt{{s_{0} \over s}} 
L_{c} \approx 0.1 {m_{\rho}^{2} \over M^{2}+ Q^{2}}
\sqrt{{E \over 1~{\rm GeV}}}\,.
\label{eq:28}
\eeq
In the numerical evaluation in (\ref{eq:28}) we used the standard 
parameters of the $\rho^{0} N$ interaction as cited in section 3 
and normal nuclear density. We see that the mass shift (\ref{eq:1}) amounts to a 
renormalization 
\beq
k_{\gamma^{*}h} \to k_{\gamma^{*}h}(1-\eta)\, .
\label{eq:29}
\eeq 
At $E=2$ GeV  the parameter $\eta \ll 1$. On the other hand, the coherence 
length (\ref{eq:25}) is very short,
$$
L_{c}(E=2~{\rm GeV}) = 1.5~{\rm fm} \cdot
{m_{\rho^{0}}^{2} \over M^2 + Q^2 }\, ,
$$
much smaller than nuclear radii for heavy nuclei and even for the light
Be nucleus it is comparable to the nuclear radius. For this reason nuclear
effects are dominated by the attenuation of the $\rho^{0}$ due to the
diagonal $\rho^{0}N \to \rho^{0}N$ transitions and, most significantly,
by distortions due to the nuclear form factor effects.

The results for the mass spectrum are shown in figs. 1a-6a by a solid curve.
For the Be target the principal effect is the mass spectrum drops at 
$M \gsim 1$ GeV much
faster than for the free nucleon target. For the heavy Pb target the form 
factor oscillations lead to an effective splitting of the $\rho^{0}$
peak - the mass spectrum develops a minimum at $M\sim m_{\rho^{0}}$. 
At $Q^{2}=1$ GeV$^{2}$ the coherence
length becomes still smaller and the distortion of the $e^{+}e^{-}$ 
mass spectrum by form factor effects become much stronger, such that
the dip at the $\rho^{0}$ mass evolves already for the Be target.
The corresponding Argand diagrams of figs. 8a and 9a span the mass 
range $0.5 < M < 1$ GeV and exhibit a structure more complex than 
a single resonance loop. Our findings for real photoproduction off
Fe and Pb targets are similar to those obtained in ref. \cite{K1} for
the target mass number $A=50$ and $A=200$ in the approximation 
of uniform nuclear density.

Notice that nowhere the splitting of the $\rho^{0}$ mass spectrum 
looks as a superposition of two Breit-Wigner peaks with the in-vacuum 
$\rho^{0}$ mass and the in-medium mass shifted by $\approx 50$ MeV
as cited in section 3. 
The weak impact of the in-medium shift (\ref{eq:1}) on the $\rho^{0}$
splitting is obvious from dashed curves in figs. 1a-6a which 
show the mass spectrum obtained when 
the mass shift (\ref{eq:1}) is neglected, i.e., putting $Re/Im = 0$. 
The distortions of the mass spectrum  change little, as a matter of 
fact the splitting of the $\rho^{0}$ peak is even  enhanced somewhat in 
conformity to the rescaling (\ref{eq:29}). Despite the slow rise of
the parameter $\eta$ with
energy, in the range $E=$2-20 GeV of the interest for experiments at
the Jefferson laboratory $\eta \ll 1$. Furthermore, at higher energies 
the contribution from the in-medium decays decreases and the overall
effect of the mass shift becomes still weaker.

\subsection{Higher energies: opening of the $\rho'(2S)$ channel.}

The new feature of high energy photoproduction is opening of the 
$\rho'(2S)$ channel and coupled-channel effects. For the comparison
purposes, we show by the thick solid curve in figs. 1b-6b the mass 
spectrum for the free nucleon target. The full coupled-channel 
results for nuclear targets comprise the attenuation, nuclear
form factor effects, the effects of $\rho^{0}(1S)
 \leftrightarrow \rho'(2S)$ transitions
including the multistep transitions and the resonance mass shift 
due to $Re/Im \neq 0$. The relative importance of these effects 
depends on the mass region, energy and the target mass number.

The importance of the off-diagonal transitions in the target nucleus
can be judged from the
comparison of the full coupled-channel results (thin solid curve) 
with those from the diagonal approximation (dotted curve) in which
off-diagonal rescatterings are switched off, i.e., $\nu=0$, and 
only the direct $\gamma^{*}\rightarrow \rho^{0}(1S)$ and 
$\gamma^{*}\rightarrow \rho'(2S)$ transitions followed by 
elastic $\rho^{0} N$ and $\rho'(2S) N$ rescatterings are included.
One can see that in the $\rho^{0}$-meson mass region $0.5<M\lsim 1$ GeV
the effect of the off-diagonal rescatterings for both the values of 
$Q^{2}$ turns out to be small. Indeed, the direct $\gamma^{*}\to \rho^{0}(1S)$
transition is strong whereas the both transitions in the off-diagonal
sequence $\gamma^{*}\to \rho'(2S) \to \rho^{0}(1S)$ are weak, see the description
of diffraction matrix in section 3. Hereafter in our discussion
of the off-diagonal effect we focus on the $\rho'(2S)$ contribution
because the contribution of intermediate states heavier than 
$\rho'(2S)$-meson is still smaller.

Our results for real photoproduction,
$Q^{2}=0$, in the $\rho^{0}(1S)$ 
mass region $^{56}$Fe  and $^{207}$Pb targets are close
to that of ref. \cite{K1} obtained for real photons for
the nucleus mass numbers $A=50$ and $A=200$ in the
approximation of uniform nuclear density.

At $Q^{2}=1$ GeV$^{2}$ the coherence length is short also for
$E=5$ GeV and distortions 
of the shape of the $e^{+}e^{-}$ mass spectrum in the
$\rho^{0}(1S)$-meson mass region by the nuclear form factor 
as described in subsection 4.5 persist for $E=5$ GeV as well.
The dominance of the form factor effects is obvious 
from the fact that these distortions change little from 
the full coupled-channel to diagonal case.

At higher masses the coupled channel effects become more
important. Here the direct $\gamma^{*}\to \rho'(2S)$
transition is weak and there is a strong interference
with the off-diagonal 
sequential transition $\gamma^{*}\to \rho^{0}(1S) \to \rho'(2S)$ 
which contains a comparably weak transition $\rho^{0}(1S) \to \rho'(2S)$.
As a matter of fact, for heavy target nuclei the sequential mechanism 
$\gamma^{*}\rightarrow\rho^{0}(1S)\rightarrow \rho'(2S)$ is found
to dominate over the direct mechanism 
$\gamma^{*}\rightarrow \rho'(2S)$, which can be seen as follows.
In figs.~1-6 we show by the dashed and long-dashed curve the 
pure $\rho^{0}(1S)$ and the $\rho'(2S)$ contributions evaluated in the full
coupled-channel approach, $\nu=2$. The dot-dashed curve shows
the pure $\rho'(2S)$ contribution evaluated in the diagonal
approximation, $\nu=0$, and in this approximation the $\rho'(2S)$ 
signal is much weaker than in the coupled-channel case in which
the $\rho'(2S)$ is fed by sequential transitions.
This is a very interesting example when the $\rho'(2S)$ production
off heavy nuclei opens a possibility for extracting 
the matrix element $\langle \rho'(2S)|\hat\sigma|\rho^{0}(1S)\rangle$ 
from the experimental data on the $e^{+}e^{-}$ mass spectrum in the $\rho'(2S)$ 
mass region. The measurement of this matrix element could give 
a unique information on the overlap of the $\rho^{0}(1S)$ and $\rho'(2S)$ 
wave functions.

The $\rho^{0}(1S)$-$\rho'(2S)$ interference changes from destructive at masses
$M$ below the $\rho'(2S)$ peak to constructive at, and above, the $\rho'(2S)$
peak, reflecting the mass dependence of the relative phase of the 
$\rho^{0}(1S) \to e^+e^-$ and $\rho'(2S)\to e^+e^-$ Breit-Wigner
amplitudes: i) the mass spectrum develops a dip in between the
$\rho^{0}(1S)$ and $\rho'(2S)$ resonance peaks, where the coupled-channel, 
thin solid, curve goes below the thin solid curve for the 
pure $\rho^{0}(1S)$ contribution,
ii) at, and beyond, the $\rho'(2S)$ peak the coupled-channel, thin
solid curve goes well above the long-dashed curve for the pure
$\rho'(2S)$ contribution, which testifies to a substantial contribution  
form the large-mass tail of the $\rho^{0}(1S)$ in the $\rho'(2S)$ region.
Evidently, that makes impossible an experimental extraction of the cross section 
of $\rho'(2S)$ production in a probabilistic approach when the interference 
with the Breit-Wigner tail of $\rho^{0}(1S)$-meson is neglected.

\subsection{The form factor effects in the $\rho'(2S)$ region.}

For moderate energies and heavier nuclei, and also for larger $Q^{2}$,
we encounter the situation when the coherence length $L_{c}$ 
for the $\rho'(2S)$ mass region becomes comparable to and/or smaller
than the nuclear radius $R_{A}$. In this case the $\rho'(2S)$ signal
will be subject to distortions by the form factor effect in 
precisely the same manner as the $\rho^{0}$ signal at lower energies,
see section 4.5. For instance, in real photoproduction off
$^{56}Fe$ target at $E=5$ GeV the $\rho'(2S)$ peak splits into 
two bumps with the dip at $M\approx m_{\rho'}$, see fig. 3b. 
The point that this dip is connected with the  form factor 
effects is evident from the shift of the dip towards the smaller 
values of $M$, and the development of the secondary dip, 
with the increase of $Q^{2}$, compare figs. 3b and 4b.
The results for heavy target, $^{207}Pb$, fig. 5b to fig. 5c 
to fig 5d show clearly how
the dip-bump structure moves to higher masses $M$ with the 
increase of the coherence length $L_{c}$ with rising energy $E$.

A comparison of Argand diagrams for the free nucleon target, fig.~7,
and nuclear targets, figs.~8 and 9, shows that the resonance 
loop corresponding to the 
$\rho'(2S)$-meson becomes 
well visible only at $E\gsim 10$ GeV. However, even in this energy region the 
contribution to the amplitude from the Breit-Wigner tail of $\rho^{0}(1S)$ meson
cannot be neglected as compared with that from the $\rho'(2S)$ meson.
\section{Conclusions}

We have performed a coupled-channel analysis of 
nuclear-medium modification of the $\rho^{0}(1S)$- and $\rho'(2S)$-mesons in the coherent
$\gamma^{*}\rightarrow \rho^{0}(1S)A, \rho'(2S) A\rightarrow e^{+}e^{-}A$ reaction
in the kinematic region $E\sim 2-20$ GeV and $Q^{2}\lsim 1$ GeV$^{2}$.
Our findings on the interplay of the inside and outside decays,
 in-medium modifications of the inside component and the coupled-channel
effects can be summarized as follows:
\\
(i) In the $\rho^{0}(1S)$ meson mass region $0.5\lsim M\lsim 1$ GeV the effect
of the off-diagonal rescatterings is small.
For heavy nuclei in this mass region our results agree with
those obtained in ref. \cite{K1} within the one-channel approximation.\\
\noindent
(ii) The off-diagonal rescatterings become important at $M\gsim 1$ GeV.
The main off-diagonal contribution is the $\rho'(2S)$ production
through the sequential mechanism 
$\gamma^{*}\rightarrow\rho^{0}(1S)\rightarrow \rho'(2S)$.
This mechanism dominates the cross section of $\rho'(2S)$ production 
for heavy nuclei.\\
\noindent
(iii) At low energies $E\lsim 5$ GeV the shapes of the $\rho^{0}(1S)$ and $\rho'(2S)$ 
resonances are strongly
affected by the nuclear effects connected with the interference
interplay of the resonance decays inside and outside the target
nucleus and form factor effect connected with the longitudinal momentum 
transfer.\\ 
\noindent
(iv) The $\rho'(2S)$ resonance is seen well only at $E\gsim 10$ GeV.
Even at high energy its shape is affected by the interference with
the Breit-Wigner tail of $\rho^{0}(1S)$-meson which must be
included properly in an analysis of the experimental
data on the $e^{+}e^{-}$ mass spectrum in the $\rho'(2S)$ mass region.\\

In our analysis we focused on the dilepton 
decay mode. Evidently, very similar effects must be observed in measurements of 
the $\rho^{0}(1S)$ and $\rho'(2S)$ 
production through the $\pi\pi$ decay mode.
The major difference from the dilepton mode would come from to the 
final-state interaction of the $\pi\pi$ system, which would reduce
the relative contribution of the inside component 
as compared with that for the $e^{+}e^{-}$ mode. 
For this reason a comparison
of the mass spectra for these two cases would be of great interest.
It is especially interesting at low energies $E\sim 2-5$ GeV 
for heavy nuclei when the inside component is large for the
$e^{+}e^{-}$ decay mode and will be strongly suppressed by the final-state
absorption for the $\pi\pi$ mode.
The comparative theoretical analysis of the coherent $\rho^{0}(1S)$ and $\rho'(2S)$
photo- and electroproduction for the $e^{+}e^{-}$ and $\pi\pi$ 
decay modes is now in progress.
\\
\vspace{.5cm}

\section*{Acknowledgments}

The work of BGZ was partially supported by the grants INTAS
96-0597 and DFG 436RUS17/11/99.


\newpage
{\large\bf Figure captions:}
\begin{itemize}

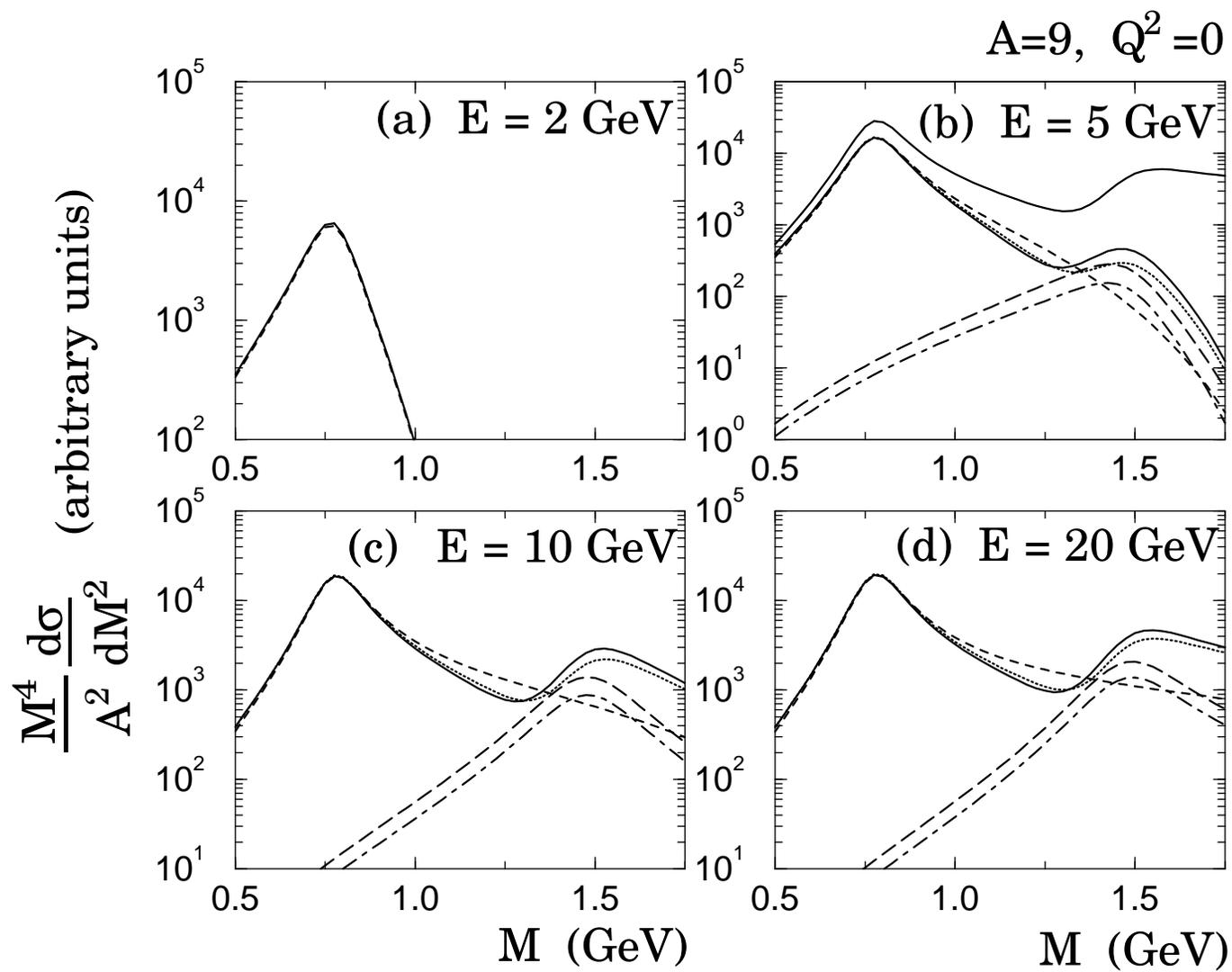
\item[Fig.~1]
~-
The $e^+e^-,\mu^+\mu^-$ rescaled mass spectrum  
(in arbitrary units) 
for coherent real photoproduction, $Q^{2}=0$,  off the $^{9}$Be nucleus
for the incident beam energy $E= 2, 5, 10$ and 20 GeV. 
The legend of curves in box {\sl (a)}: The solid curve is a prediction of the Glauber
approximation with mass shift (\ref{eq:10}) from the reggeon amplitude 
{\protect (\ref{eq:23})} included, the dashed line shows the results
obtained neglecting the mass shift.
The legend of curves in box {\sl (b)}: 
The thick solid curve is the spectrum for the nucleon 
target, all other curves are for nuclear target. The thin solid curve 
is the full coupled-channel calculation, $\nu =2$ in expansion 
{\protect (\ref{eq:10})}. The dotted curve is 
a result from the diagonal approximation, $\nu=0$ in expansion 
{\protect (\ref{eq:10})}, the $\rho^{0}(1S)$-$\rho'(2S)$ interference in the
$e^{+}e^{-}$ and/or  $\mu^+\mu^-$ decay channels included. 
The dashed curve shows the pure $\rho^{0}$ signal in the coupled-channel
calculation, $\nu=2$. The long-dashed curve is the $\rho'(2S)$ signal in 
the coupled-channel calculation, $\nu=2$. The dot-dashed curve is 
the $\rho'(2S)$ signal in the diagonal approximation, $\nu=2$. Boxes {\sl (c)}
and {\sl (d)} are the same as box {\sl (b)} but for $E=10$ and 20 GeV,
respectively, and with omission of mass spectrum for the free nucleon 
target.

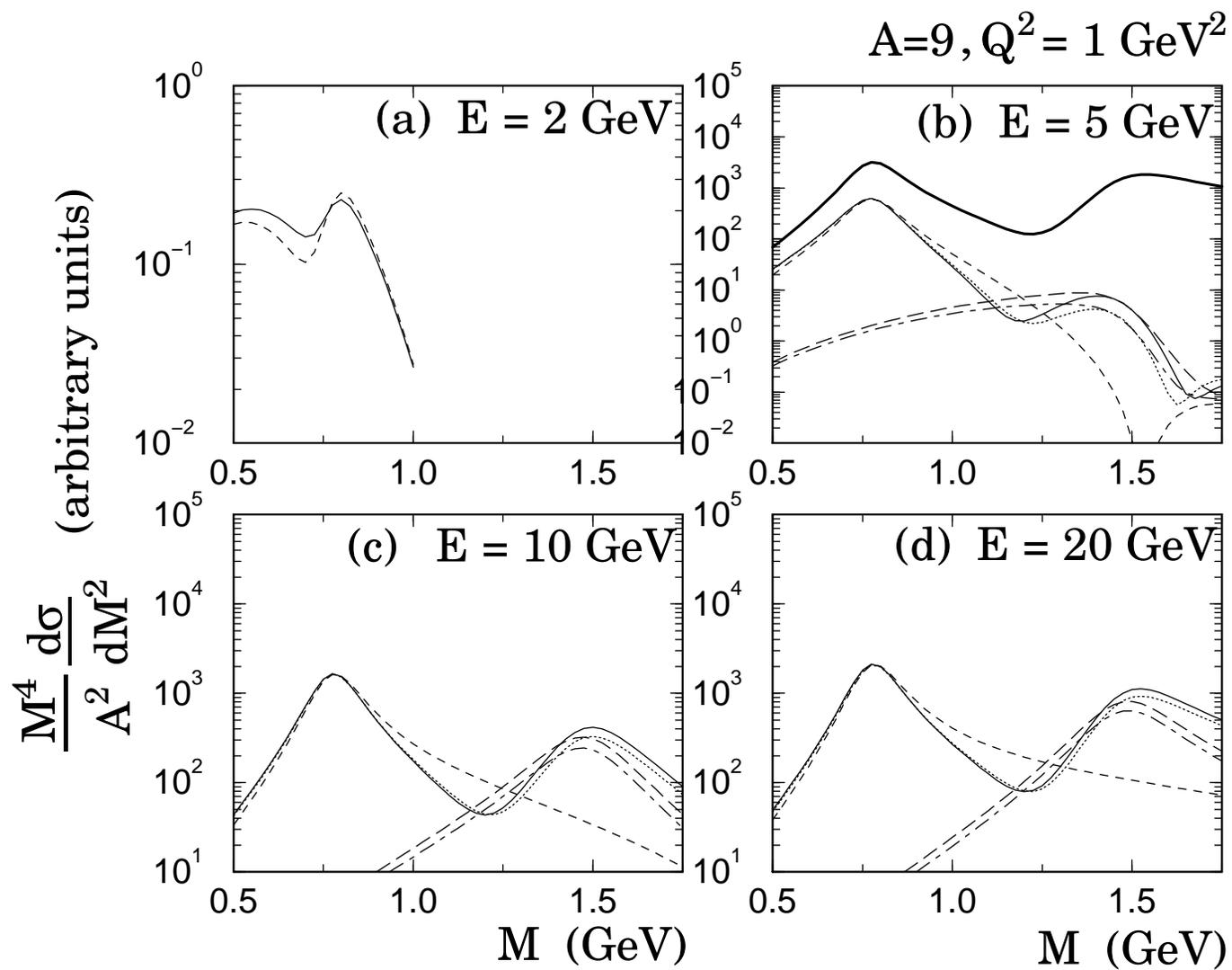
\item[Fig.~2]
~-
The same as Fig.~1 but for virtual photoproduction, $Q^{2}=1$ GeV$^{2}$.

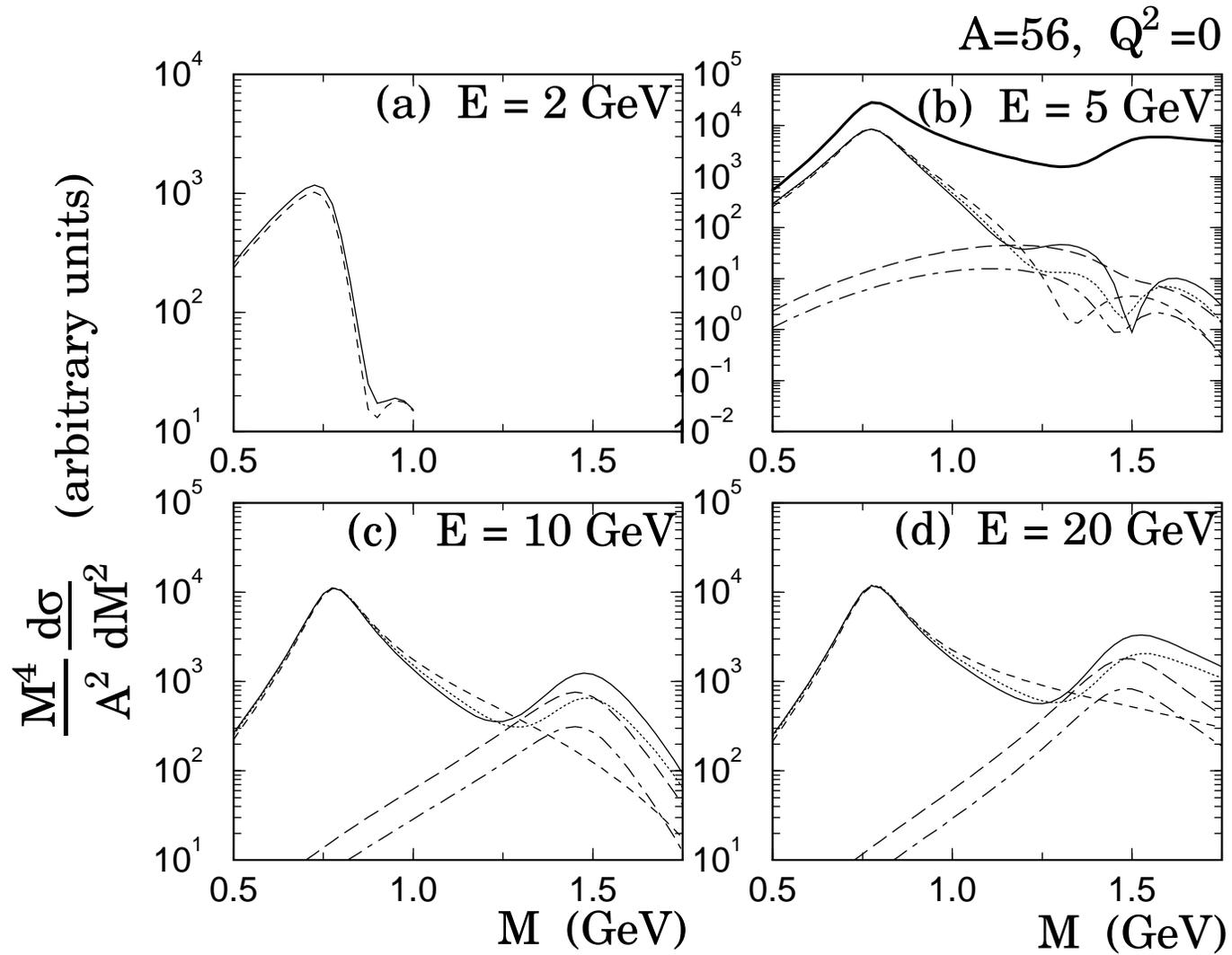
\item[Fig.~3]
~-
The same as Fig.~1 but for the target nucleus $^{56}$Fe.

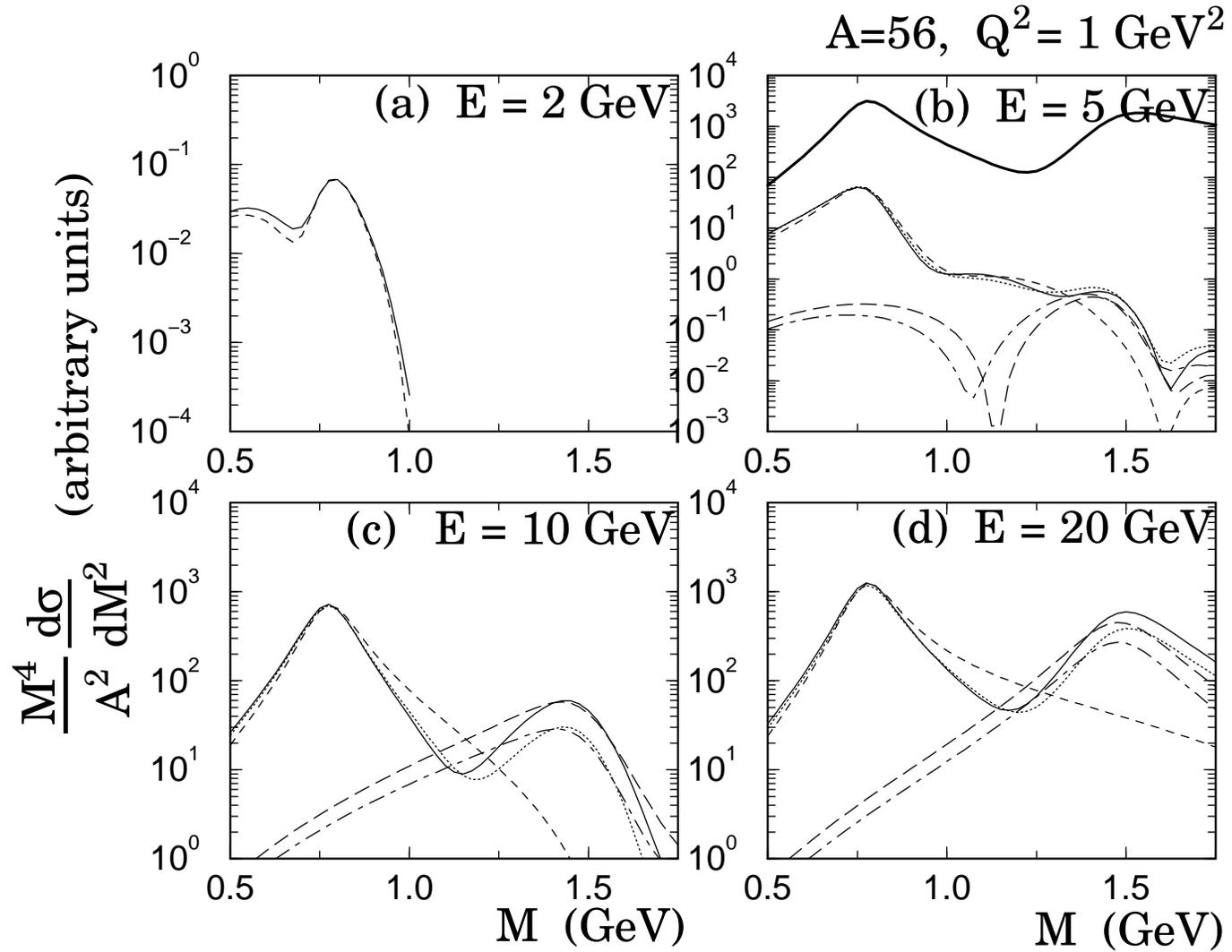
\item[Fig.~4]
~-
The same as Fig.~2 but for the target nucleus $^{56}$Fe.

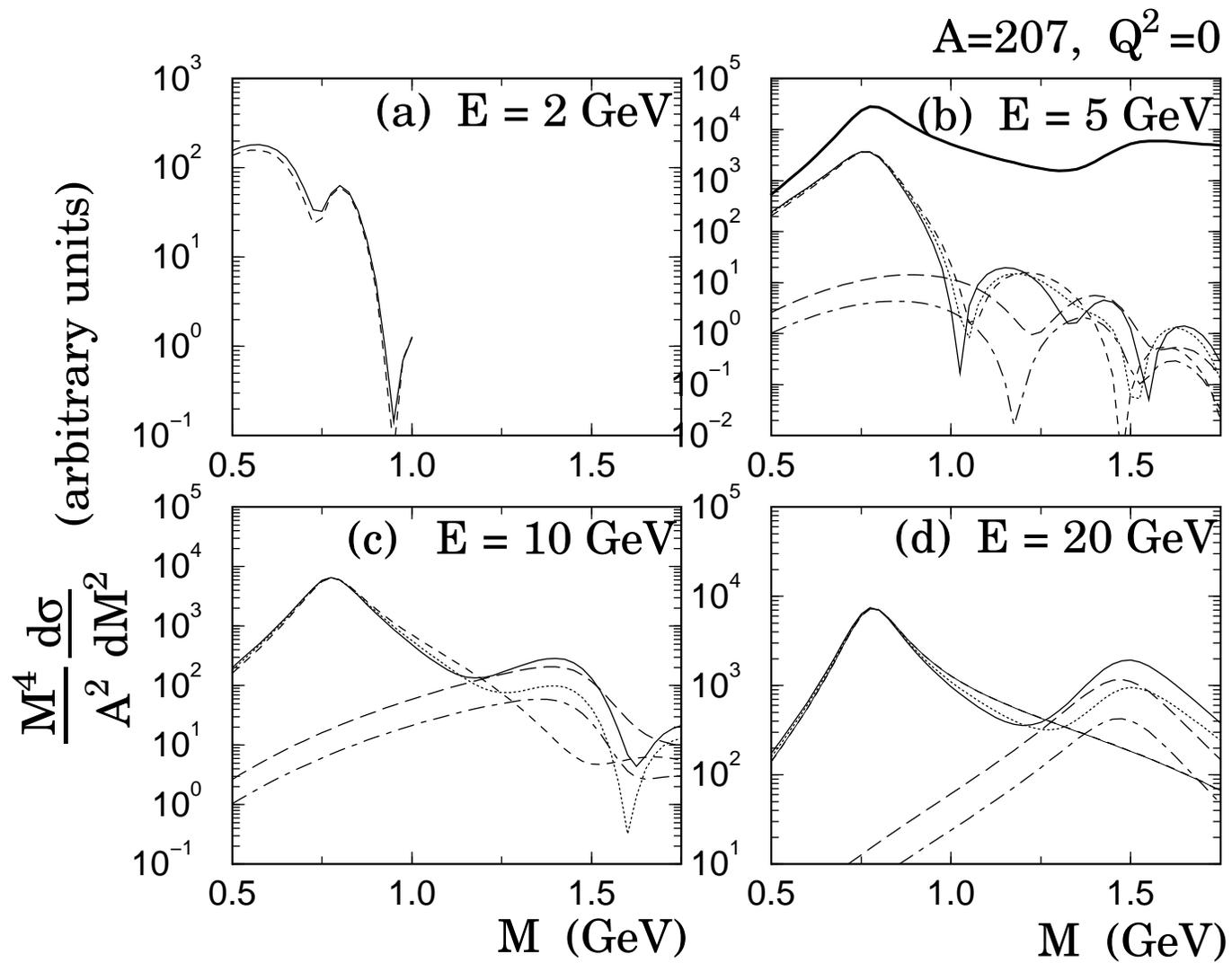
\item[Fig.~5]
~-
The same as Fig.~1 but for the target nucleus $^{207}$Pb.
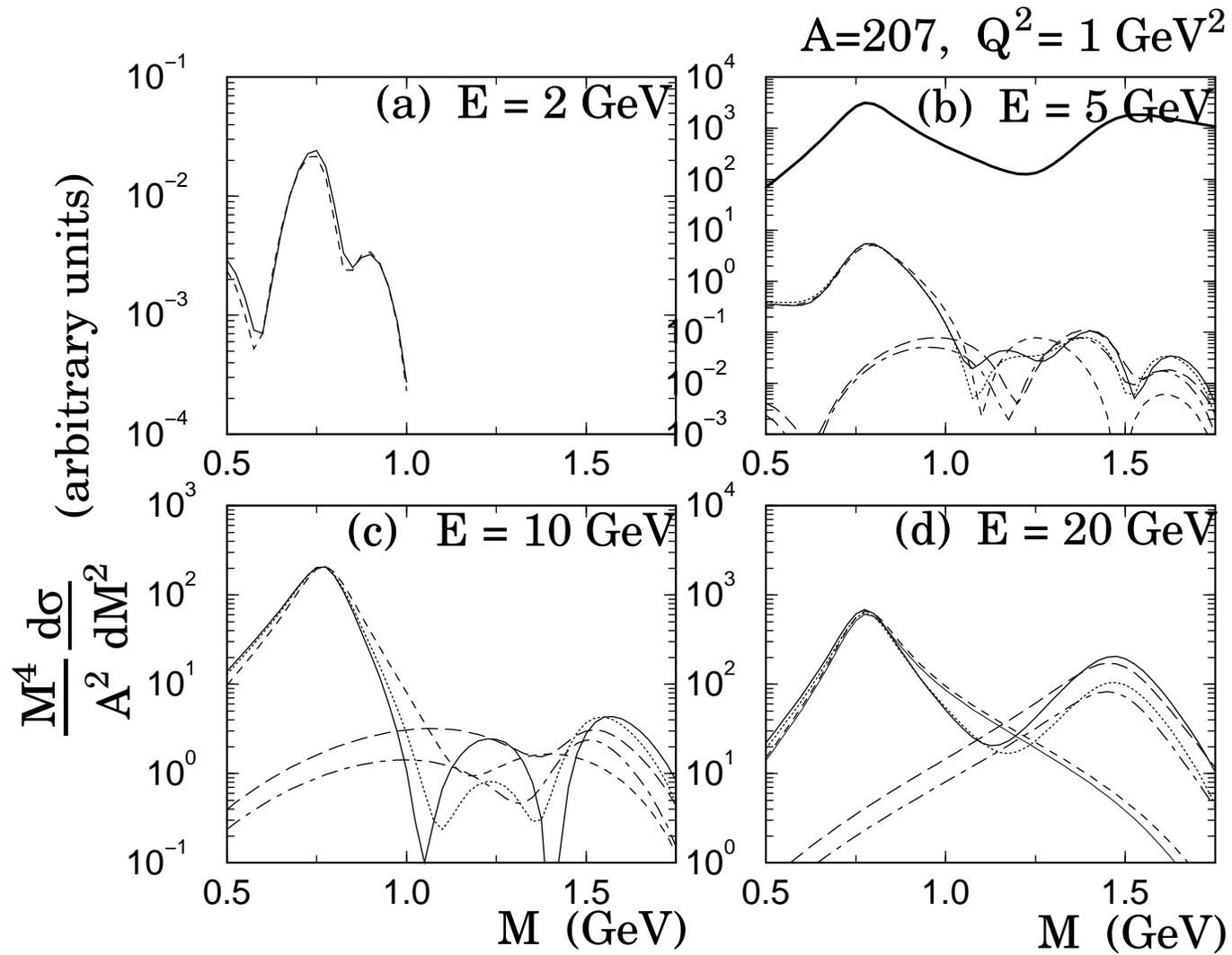
\item[Fig.~6]
~-
The same as Fig.~2 but for the target nucleus $^{207}$Pb.

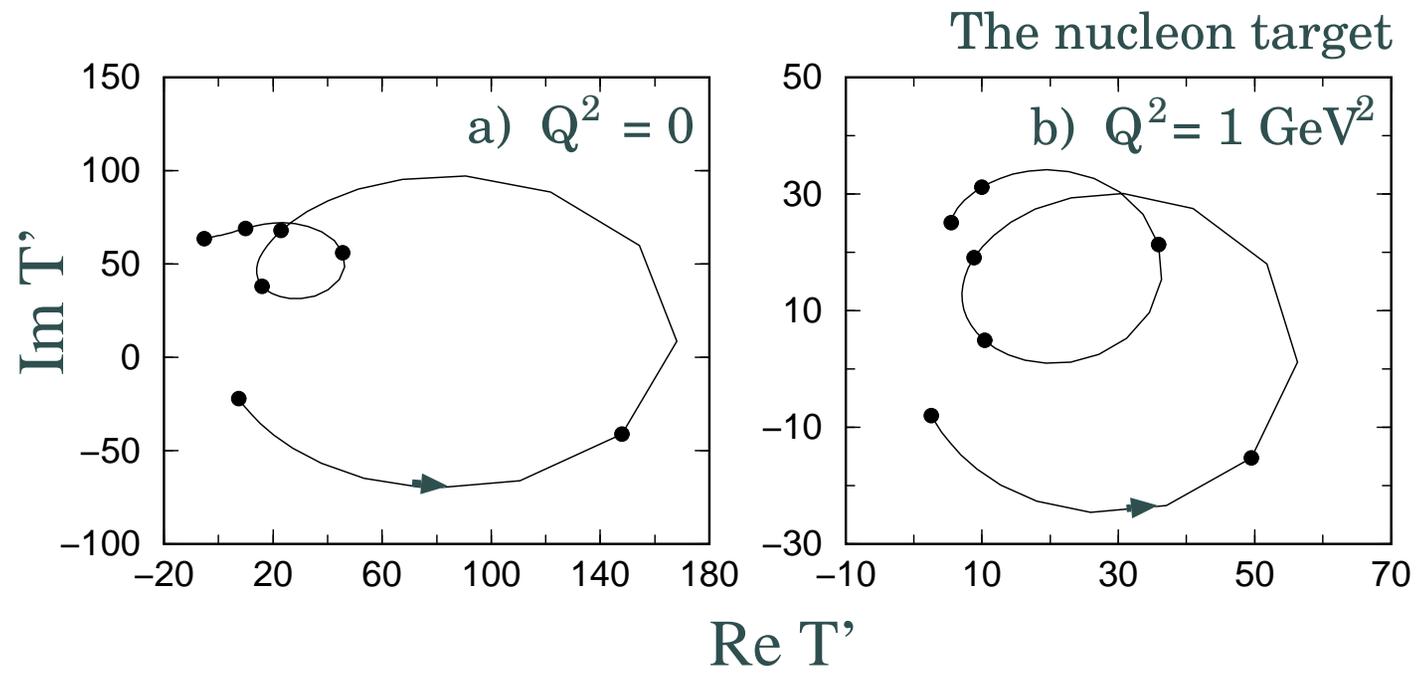
\item[Fig.~7]
~-
The Argand plots for the scaled amplitude 
$T'=M^{2}T(E,M)$ for {\sl (a)} real, $Q^{2}=0$, and
{\sl (b) } virtual, $Q^{2}=1$ GeV$^{2}$, photoproduction off
the free nucleon target. 

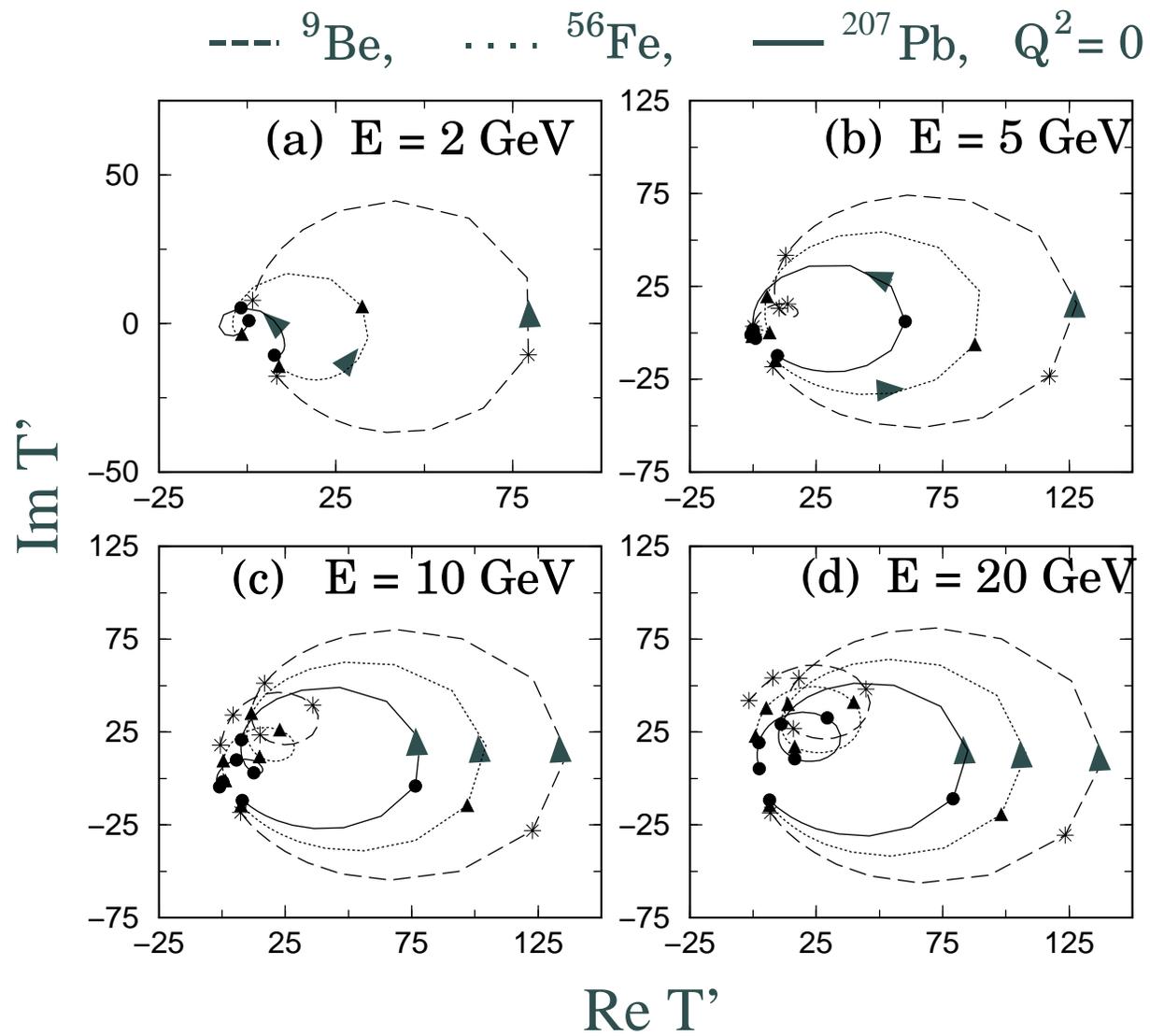
\item[Fig.~8]
~-
~-
The Argand plots for the scaled amplitude 
{\protect (\ref{eq:26})}
at $Q^{2}=0$. The solid, dotted, and dashed curves show
the results for the target nucleus
$^{207}$Pb, $^{56}$Fe, and $^{9}$Be, respectively.
The curves are given in arbitrary units. The mass intervals
are 0.5-1 GeV for $E=2$ GeV, and 0.5-2 GeV for $E=5,$ 10, and 20 GeV.
The spacing of mass points along the curves is 0.25 GeV, the 
arrows show the direction of the increasing mass.

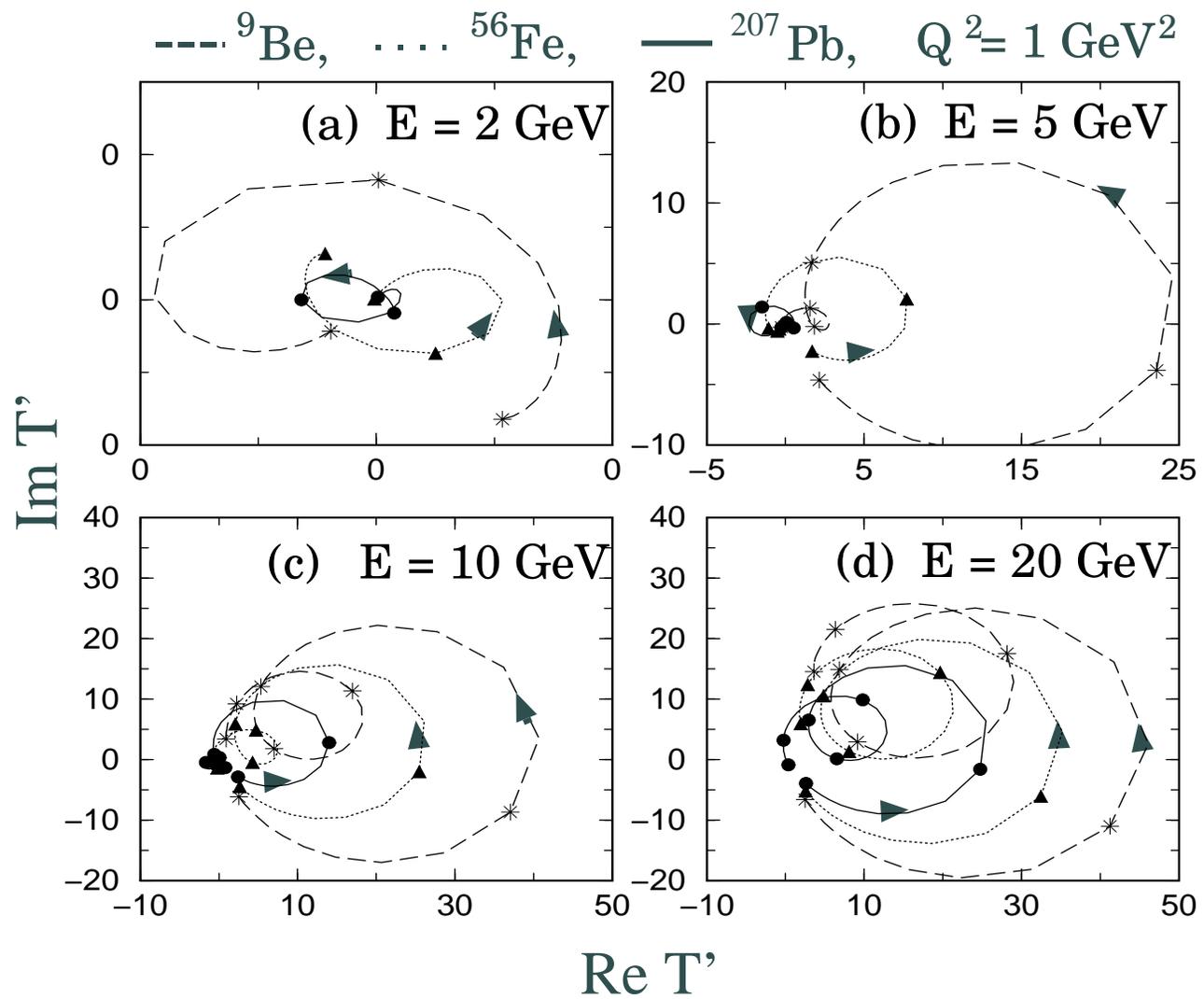
\item[Fig.~9]
~-
The same as Fig.~8 but for $Q^{2}=1$ GeV$^{2}$.

\end{itemize}

\end{document}